\documentclass[12pt]{article}
\usepackage{amsmath}
\usepackage{amssymb}
\usepackage{array}
\usepackage{geometry}
\usepackage{graphicx}
\usepackage{enumerate}
\usepackage[small, compact] {titlesec}
\numberwithin{equation}{section}

\begin{document}

\begin{center}
{\bf \large Fusion yield: Guderley model and Tsallis statistics}\\
\bigskip
{\bf H.J. Haubold$^1$$^,$$^2$ and D. Kumar$^2$}\\
\small{$^1$Office for Outer Space Affairs, United Nations, \\
Vienna International Centre, P.O. Box 500, A-1400 Vienna, Austria\\
$^2$Centre for Mathematical Sciences Pala Campus\\
Arunapuram P.O., Palai, Kerala  686 574, India}
\end{center}
\begin{center}
{\bf Abstract} \\
\end{center}

{ \small
The reaction rate probability integral is extended from Maxwell-Boltzmann approach to a more general approach by using the pathway model introduced by Mathai [Mathai A.M.:2005, A pathway to matrix-variate gamma and normal densities, {\it Linear Algebra and Its Applications}, {\bf 396}, 317-328].  The extended thermonuclear reaction rate is obtained in closed form via a Meijer's $G$-function and the so obtained $G$-function is represented as a solution of a homogeneous linear differential equation.  A physical model for the hydrodynamical process in a fusion plasma compressed and laser-driven spherical shock wave is used for evaluating the fusion energy integral by integrating the extended thermonuclear reaction rate integral over the temperature.  The result obtained is compared with the standard fusion yield obtained by Haubold and John in 1981.[Haubold, H.J. and John, R.W.:1981, Analytical representation of the thermonuclear reaction rate and fusion energy production in a spherical plasma shock wave, {\it Plasma Physics}, {\bf 23}, 399-411].  An interpretation for the pathway parameter is also given.\\

}\\
{\section{\bf Introduction}}

 Energy production in the Sun and other stars is mainly due to thermonuclear reactions taking place in them \cite{fowler1984,davis2003,bogoyavlensky1985}.  Hence the nuclear reactions in the laboratory situation for energy production has got more importance especially when the reactions are laser-induced fusion reactions \cite{brueckner1974,atzenimeyertervehn2009}.  The fusion reactions are controlled by the thermonuclear reaction rates under specific theoretical assumptions and experimental inputs.  The major facts which are required for the calculation of the reaction rates are (1) the quantum-mechanical cross section of the reaction and (2) the statistical mechanics velocity distribution of the reacting particles.  A systematic study of reaction rate theory has been done by many authors \cite{mathaihaubold1988,hauboldmathai1998,hauboldjohn1978}.  In the literature it is assumed that the physical parameters in the solar interior (particularly density and temperature) lead to an equilibrium velocity distribution for fusion plasma particles.  This distribution of particles is assumed to be Maxwell-Boltzmannian in almost all cases in stellar physics and cosmology \cite{hauboldkumar2008}.  A small deviation to the distribution of the reacting particles results in a change in the reaction rates.  It was observed that the distribution function in the solar interiors cannot be much different from Maxwell-Boltzmannian \cite{coradduetal1999, deglinnocenti1998}.\\

The dynamics of an imploding spherical shock front in inertial confinement fusion was first investigated by Guderley using a self-similar form of the fluid equations \cite{guderley1942}. This analysis was continued by others \cite{lazarus1981,hafner1988} and has been reviewed extensively by Zel'dovich and Raizer \cite{zeldovichraizer2002} and Atzeni and Meyer-Ter-Vehn \cite{atzenimeyertervehn2009}. Brueckner and Jorna \cite{brueckner1974}, Haubold and John \cite{hauboldjohn1981}, and Rygg {cite{rygg2006} calculated analytic approximations to the respective fusion energy yields. In such calculations, Maxwell-Boltzmannian velocity distributions are assumed for reacting particles but most likely this assumption is violated due to non-locality of the constituent particles and the convergence of the fuel at velocities comparable to the typical ion thermal velocity \cite{rygg2006}.\\

The spatio-temporal dynamics of solar activity has been investigated by studying the main solar oscillations and the time evolution of the basic periods \cite{vecchiocarbone2009}. Period length variations could improve knowledge of the relationship between observed solar activity due to variable phenomena generated by solar dynamo models or even in the gravitationally stabilized solar fusion reactor \cite{wolff2002, wolff2007, wolff2009, grandpierre2010} due to effects of a deep mixed shell on solar g-modes, p-modes, and neutrino flux.\\ 

 Tsallis, in 1988, suggested a generalization of Boltzmann-Gibbs statistical mechanics with a $q$-exponential function  as distribution function and $q \rightarrow1$ recovers Boltzmann-Gibbs statistics \cite{tsallis2009,gellmantsallis2004}.  An attempt was made to extend the theory of nuclear reaction rates from the Maxwell-Boltzmann approach to the Tsallis approach by many authors \cite{saxenaetal2004,mathai2005,mathaihaubold2007, mathaihaubold2008, mathaietal2010}. In 2005, Mathai \cite{mathai2005,hauboldetal2010} discovered a pathway model which covers the Tsallis distribution and many other distribution functions as special cases.  Mathai \cite{mathai2005} mainly deals with rectangular matrix-variate distributions and the scalar case is a particular one.  The real scalar case of the pathway model has the following forms:
  \begin{equation}\label{type1pathway}
f_1(x)=c_1|x|^{\gamma}[1-a(1-\alpha)|x|^\delta]^{\frac{1}{1-\alpha}},~~
a>0,\delta>0, 1-a(1-\alpha)|x|^\delta>0, \gamma>0, \alpha<1
\end{equation}
and $f_1(x)=0$ when $1-a(1-\alpha)|x|^\delta \leq 0,$ where $\alpha$ is the pathway parameter. When
$\alpha<1$ the model is a generalized type-1 beta model in the real case.  When $\alpha>1$ we have from (\ref{type1pathway})
\begin{equation}\label{type2pathway}
f_2(x)=c_2|x|^{\gamma}[1+a(\alpha-1)|x|^\delta]^{-\frac{1}{\alpha-1}},~-\infty<x<\infty
\end{equation}
which is a generalized type-2 beta model for real $x>0$.  When
$\alpha\rightarrow1$ the above two forms will reduce to
\begin{equation}\label{gammapathway}
f_3(x)=c_3|x|^{\gamma}{\rm e}^{-a |x|^\delta},
\end{equation}
where the normalizing constants are  given by
\begin{equation}\label{constanttype1}
c_1=\frac{\delta[a(1-\alpha)]^{\frac{\gamma+1}{\delta}}\Gamma\left( \frac{\gamma+1}{\delta}+
\frac{1}{1-\alpha}+1\right)}{\Gamma\left( \frac{\gamma+1}{\delta}\right)\Gamma\left(
\frac{1}{1-\alpha}+1\right)},~ \alpha<1,
\end{equation}
\begin{equation}\label{constanttype2}
c_2=\frac{\delta[a(\alpha-1)]^{\frac{\gamma+1}{\delta}}\Gamma\left( \frac{1}{\alpha-1}\right)}{
\Gamma\left( \frac{\gamma+1}{\delta}\right)\Gamma\left( \frac{1}{\alpha-1}-\frac{\gamma+1}{\delta}\right)}
,~ \Re\left(\frac{1}{\alpha-1}-\frac{\gamma+1}{\delta} \right)>0,~\alpha>1
\end{equation}
and
\begin{equation}\label{constantgamma}
c_3=\frac{\delta (a\eta)^{\frac{\gamma+1}{\delta}}}{\Gamma(\frac{\gamma+1}{\delta})},\alpha\rightarrow1,
\end{equation}
where $\Re(.)$ denotes the real part of $(.)$.  For different special cases of the above forms see \cite{mathai2005, mathaihaubold2008}.  The pathway model can also be established from a generalized entropy of order $\alpha$ using the maximum entropy principle \cite{mathaihaubold2007,mathaihaubold2007a}.  Haubold and Kumar in 2008 \cite{hauboldkumar2008} extended the  reaction rate theory to a general form covering the Tsallis reaction rates and established the closed form representation of the extended thermonuclear reaction rate in terms of generalized special functions, mainly in Meijer's $G$-function and $H$-function \cite{mathai1993, mathaisaxena1973, mathaisaxena1978, mathaietal2010}.\\

In this paper we develop a complete theory for the representation of the nuclear fusion yield integral by using the closed form representation of the extended thermonuclear reaction rate.  Brueckner and Jorna \cite{brueckner1974} give a physical foundation of laser-induced nucleosynthesis. Instead of using difficult physical descriptions of laser-induced fusion, a simple physical model of  Guderley \cite{guderley1942} has been used and fusion energy yield integral has been evaluated in closed form via a $G$-function. From the literature it is observed that Maxwell-Boltzmann distribution function is used in the model calculations concerning fusion reactions in strong plasma shock waves. But we use pathway model \cite{mathai2005,mathaihaubold2008} which has a more general distribution function than Maxwell-Boltzmann distribution, for the calculations.  As the  pathway parameter $\alpha \rightarrow 1$ the Maxwell-Boltzmann distribution is retrieved.\\

The paper is organized as follow:  In the next section we extend the thermonuclear reaction rate using the pathway density function and present the result in closed form in terms of a $G$-function.  A homogeneous linear differential equation which satisfies the $G$-function is also obtained.  In section 3 we illustrate the model formation by using the Guderley model \cite{guderley1942} for the compressed fusion plasma and heated by a strong spherical shock wave used by Haubold and John \cite{hauboldjohn1981} and Brueckner and Jorna \cite{brueckner1974}.  Section 4 gives the exact evaluation of the fusion yield integral by the extended reaction rate.  A comparison of the new result obtained with the result already obtained by Haubold and John \cite{hauboldjohn1981} is done in section 5. A physical interpretation for the pathway parameter in the newly obtained result is also given. Concluding remarks are added in section 6.
{\section{\bf Evaluation of the extended thermonuclear reaction rate in closed form}}

We consider the reaction between two interacting particles of certain type $i$ and $j$, then the reaction rate $r_{ij}$ is the product of the number densities $n_i$ and $n_j$ and the reaction probability $\langle\sigma v\rangle_{ij}$ between the particles which is given by
\begin{equation}\label{reactionrate}
r_{ij}=(1-\frac{1}{2}\delta_{ij})N_i N_j \langle\sigma v\rangle_{ij}.
\end{equation}
The bracketed quantity $\langle\sigma v\rangle_{ij}$ is the probability per unit time that two particles $i$ and $j$ confined to a unit volume will react with each other.  It is the statistical average of the reaction cross section $\sigma(v)$ over the normalized distribution function $n(v)$ of the relative velocity $v$ of the center of mass system:
\begin{equation}\label{reactionprobability}
\langle\sigma v\rangle_{ij}=\int_0^\infty
{\rm d}v~ n(v) \sigma(v) v.
\end{equation}

When we consider the cross-section $\sigma(E)$ for low-energy nuclear reactions far from any resonance, we obtain the expression for the cross-section of the low-energy non-resonant nuclear reactions as \cite{hauboldjohn1978, fowler1984}
\begin{equation}\label{reactioncrosssection}
\sigma(E)=\frac{S(E)}{E}{\rm e}^{-2\pi \eta(E)},~~\left( \frac{E}{B}\ll 1 \right)
\end{equation}
where $E$ is the relative kinetic energy between the particles, $B$ the nuclear barrier height and ${\rm e}^{-2\pi \eta(E)}$ is usually called the Gamow factor, gives the probability that the particles penetrate the potential wall.  The Sommerfeld parameter $\eta(E)$ is given by
\begin{equation}\label{sommerfeldparameter}
\eta(E)=\left( \frac{\mu}{2} \right)^{\frac{1}{2}}\frac{Z_i Z_j e^2}{\hbar E^{\frac{1}{2}}},
\end{equation}
where $Z_i$ and $Z_j$ are the atomic numbers of the nuclei $i$ and $j$, $e$ is the quantum of electric charge, $\hbar$ is the Planck's quantum of action. The cross-section factor $S(E)$ entering the cross-section is a slowly varying function of energy over a limited energy range and can be characterized depending on the nuclear reaction. Usually when $S$ is a constant one takes it as $S_0$. S(E) contains the constants intrinsic to the nuclear reactions under consideration.  Its energy dependence may be represented by a Maclaurin series expansion up to the second order in the kinetic energy $E$ as following \cite{hauboldjohn1978, hauboldjohn1981}:
\begin{equation}\label{crosssectionfactor}
S(E) \approx  S(0) + \frac{{\rm d}S(0)}{{\rm d}E}E +\frac{1}{2}
 \frac{{\rm d}^2S(0)}{{\rm d}E^2}E^2 =\sum_{\nu=0}^{2}\frac{S^{(\nu)}(0)}{\nu !}E^\nu.
 \end{equation}
Due to the strong dependence of the cross-section $\sigma(E)$ on the relative kinetic energy $E$ at the temperature of thermonuclear burning, the essential contribution to the thermonuclear reaction comes from the collisions between particles with energies more than the average thermal energy ($\frac{3kT}{2}$).  Hence the reaction probability depends on the reacting particles with energies greater than the mean thermal energy.\\

Usually, the thermonuclear fusion plasma is assumed to be in thermodynamical equilibrium with regard to their velocities (not with regard to their mass abundance).  From the tabulations of the nuclear reaction rates by Fowler et al. \cite{fowleretal1967}, we can infer that the distribution of the relative velocities of the reacting particles always remains Maxwell-Boltzmannian.  In case of non-degenerate and non-relativistic gas the distribution function of the relative velocities of the nuclei is Maxwell-Boltzmannian.
\begin{equation}\label{Maxwelboltzman}
f_{MBD}(E)= 2\pi \left( \frac{1}{\pi kT} \right)^{\frac{3}{2}}
e^{-\frac{E}{kT}}\sqrt{E} {\rm d}E.
\end{equation}

By using the Maxwell-Boltzmann distribution function (\ref{Maxwelboltzman}), the non-resonant low-energy cross-section (\ref{reactioncrosssection}) and the expression for the cross-section factor given in (\ref{crosssectionfactor}), the reaction probability in (\ref{reactionprobability}) becomes
\begin{equation}
\langle\sigma v\rangle_{ij}=\left( \frac{8}{\pi \mu} \right)^{\frac{1}{2}}\left( \frac{1}{kT} \right)^{\frac{3}{2}}\sum_{\nu=0}^{2}\frac{S^{(\nu)}(0)}{\nu !}\int_0^\infty E^\nu{\rm e}^{-\frac {E}{kT}-2\pi \left( \frac{\mu}{2} \right)^{\frac{1}{2}}\frac{Z_i Z_j e^2}{\hbar E^{\frac{1}{2}}}}{\rm d}E. \\
\end{equation}
  If we assume a deviation from the Maxwell-Boltzmann distribution function or a more general distribution function, then we can consider a distribution function $f_{MPD}(
  E)$ obtained by the pathway model of Mathai \cite{mathai2005,mathaihaubold2007} given by
  \begin{equation}\label{mathaipathway}
  f_{MPD}(E)= 2\pi  \frac{\sqrt{E}}{(\pi kT)^{\frac{3}{2}}} \left[1+(\alpha-1)\frac{E}{kT}\right]^{-\frac{1}{\alpha-1}}{\rm d}E.
  \end{equation}
  As $\alpha \rightarrow1$ in (\ref{mathaipathway}) we obtain the Maxwell-Boltzmann distribution function given in (\ref{Maxwelboltzman}).  Substituting Mathai's pathway distribution (\ref{mathaipathway}) instead of Maxwell-Boltzmann distribution (\ref{Maxwelboltzman}) in (\ref{reactionprobability}) we get,
  \begin{equation}\label{reactionprobabilitymathai}
\langle\sigma v\rangle_{ij}=\left( \frac{8}{\pi \mu} \right)^{\frac{1}{2}}\left( \frac{1}{kT} \right)^{\frac{3}{2}}\sum_{\nu=0}^{2}\frac{S^{(\nu)}(0)}{\nu !}\int_0^\infty E^\nu \left[1+(\alpha-1)\frac{E}{kT}\right]^{-\frac{1}{\alpha-1}}{\rm e}^{-2\pi \left( \frac{\mu}{2} \right)^{\frac{1}{2}}\frac{Z_i Z_j e^2}{\hbar E^{\frac{1}{2}}}}{\rm d}E. \\
\end{equation}
For the convenience in evaluating the above integral we give a substitution $y=\frac{E}{kT}$ and $x=2\pi \left( \frac{\mu}{2kT} \right)^{\frac{1}{2}}\frac{Z_i Z_j e^2}{\hbar }$.  Then the equation (\ref{reactionprobabilitymathai}) becomes
\begin{equation}\label{reactionprobabilitysubstituted}
\langle\sigma v\rangle_{ij}=\left( \frac{8}{\pi \mu} \right)^{\frac{1}{2}}\sum_{\nu=0}^{2}
\left( \frac{1}{kT} \right)^{-\nu+\frac{1}{2}}\frac{S^{(\nu)}(0)}{\nu !}
\int_0^\infty y^\nu [1+(\alpha-1)y]^{-\frac{1}{\alpha-1}}{\rm e}^{-xy^{-\frac{1}{2}}}{\rm d}y.
\end{equation}
The reaction probability $\langle\sigma v\rangle_{ij}$ is the finite sum of integrals of the form
\begin{equation}\label{i1alpha}
I_{1\alpha}=\int_0^\infty y^\nu [1+(\alpha-1)y]^{-\frac{1}{\alpha-1}}{\rm e}^{-xy^{-\frac{1}{2}}}{\rm d}y.
\end{equation}
Using the Mellin convolution property the above integral can be evaluated and expressed in closed form by using generalized special functions.  The function can be expressed in terms of a Meijer's $G$-function of one variable.  $G$-function $G^{m,n}_{p,q}(z|^{a_1,a_2,\cdots,a_p}_{b_1,b_2,\cdots,b_q})$ was introduced by Meijer in 1936 as a generalization of the hypergeometric function, see \cite{mathai1993, mathaisaxena1973, mathaietal2010} for details.  For the detailed evaluation of $I_{1\alpha}$ in (\ref{i1alpha}) see \cite{hauboldkumar2008}.  Thus we obtain
\begin{equation}\label{i1alphagfunction}
I_{1\alpha}=\frac{(\pi)^{-\frac{1}{2}}}
{(\alpha-1 )^{\nu+1 }\Gamma \left
(\frac{1}{\alpha-1 } \right )}G_{1,3}^{3,1} \left(
\frac{(\alpha-1)x^2}{4} \big|^{2-\frac{1}{\alpha-1}+\nu}_
 {0, \frac{1}{2}, \nu+1} \right).
\end{equation}
According to this the extended reaction probability in (\ref{reactionprobabilitymathai}) becomes
\begin{eqnarray}
\langle\sigma v\rangle_{ij}&=&(\pi)^{-1}\left( \frac{8}{\mu} \right)^{\frac{1}{2}} \sum_{\nu=0}^{2}\left( \frac{1}{kT} \right)^{-\nu+\frac{1}{2}}\frac{S^{(\nu)}(0)}{\nu !}\frac{1}
{(\alpha-1 )^{\nu+1 }\Gamma \left
(\frac{1}{\alpha-1 } \right )}\nonumber\\
&&\times G_{1,3}^{3,1} \left[
(\alpha-1)\pi^2 \left( \frac{\mu}{2kT} \right)\left(\frac{Z_i Z_j e^2}{\hbar }\right)^2 \big|^{2-\frac{1}{\alpha-1}+\nu}_
 {0, \frac{1}{2}, \nu+1} \right]
\end{eqnarray}
The Meijer's $G$-function $G^{3,1}_{1,3}\left(z\big|^{2-\frac{1}{\alpha-1}+\nu}_
 {0, \frac{1}{2}, \nu+1}\right)$ satisfies the homogeneous linear differential equation of order three \cite{mathaisaxena1973, mathaietal2010}.
 \begin{eqnarray}\label{diffequation}
0&=& \bigg[-z \left( z \frac{{\rm d}}{{\rm d}z}-a_1+1\right)-\prod_{j=1}^{3}\left(z \frac{{\rm d}}{{\rm d}z}-b_j\right) \bigg]G^{3,1}_{1,3}\left(z\big|^{a_1=2-\frac{1}{\alpha-1}+\nu}_
 {b_1=0, b_2=\frac{1}{2}, b_3=\nu+1}\right)\nonumber\\
&=& \bigg[-z \left( z \frac{{\rm d}}{{\rm d}z}+\frac{1}{\alpha-1}-1-\nu\right)-z \frac{{\rm d}}{{\rm d}z}\left(z \frac{{\rm d}}{{\rm d}z}-\frac{1}{2}\right)\nonumber \\ &&\times \left(z \frac{{\rm d}}{{\rm d}z}-(1+\nu)\right)\bigg]G^{3,1}_{1,3}\left(z\big|^{2-\frac{1}{\alpha-1}+\nu}_
 {0, \frac{1}{2}, \nu+1}\right)\nonumber \\
&=&\bigg[-z^3\frac{{\rm d}^3}{{\rm d}z^3}-\left(\frac{3}{2}-\nu\right)z^2\frac{{\rm d}^2}{{\rm d}z^2}+\left(\frac{\nu}{2}-z\right)z \frac{{\rm d}}{{\rm d}z}\nonumber\\
&&+\left(1+\nu-\frac{1}{\alpha-1}\right)z\bigg]G^{3,1}_{1,3}
\left(z\big|^{2-\frac{1}{\alpha-1}+\nu}_
 {0, \frac{1}{2}, \nu+1}\right).
 \end{eqnarray}
 Substituting $z=\frac{(\alpha-1)x^2}{4}$ in (\ref{diffequation}) we obtain
 \begin{equation}\label{mathaidifferentialequation}
 \left[x\frac{{\rm d}^3}{{\rm d}x^3}-2 \nu \frac{{\rm d}^2}{{\rm d}x^2}+(\alpha-1)x \frac{{\rm d}}{{\rm d}x}-2[(1+\nu)(\alpha-1)-1]\right]G^{3,1}_{1,3}
\left(\frac{(\alpha-1)x^2}{4}\big|^{2-\frac{1}{\alpha-1}+\nu}_
 {0, \frac{1}{2}, \nu+1}\right)=0
 \end{equation}
 {\section{\bf Fusion yield integral in the case of Shock-compressed and heated plasma}}
 To produce a useful amount of fusion energy the plasma pellet is to be highly and efficiently compressed.  The center of the compressed pellet must be brought to the ignition temperature and density, but the rest of the compressed pellet must be left as cold as possible. Very high compression is needed to maximize the reaction rate after the pellet is ignited \cite{brueckner1974}.  A laser can be used as a hydrodynamic driver of pellet compression.  The uniform pressure applied to the pellet surface by the laser energy will produce a single converging shock \cite{daiberetal1966, brueckner1974}.  We take the Guderley model \cite{guderley1942} of the dynamics of strong spherical shocks near the center of the sphere provided the perturbation of the hydrodynamic process by the fusion reaction is ignored.  As the converging shock approaches the center of collapse, it decouples from its generating boundary. In a strong spherically convergent shock the shock position $r$ can be written as a function of time $t_0$ as
 \begin{equation}\label{shockposition}
 r=\xi (-t_0)^n,
 \end{equation}
 where we measure the time at which the shock reaches the center of convergence.  The parameter $\xi$ measures the shock strength, for the number $n$, $0<n<1$ \cite{guderley1942}.  In spherical geometry and for an ideal gas with the ratio of specific heats $\gamma=\frac{5}{3}$, the exponent $n$ has to take the value $n=0.688377$.  The shock velocity can be determined from the equation (\ref{shockposition}) as
 \begin{equation}\label{shockvelocity}
u_0=\frac{{\rm d}r}{{\rm d}t_0}=n \xi (-t_0)^{n-1}=n \xi^{\frac{1}{n}} r^{\frac{n-1}{n}}, \end{equation}
where the time $t_0$ is negative and it increases up to the center of convergence at $t_0=0$.

The scaled density, scaled temperature and scaled pressure at a radius $r$ are functions of the reduced time $-\frac{t}{t_0}$  where $-t_0$ is the time after which the shock front has reached the center of convergence.
\begin{equation}\label{frho}
\frac{\rho}{\rho_0}=f_\rho\left(-\frac{t}{t_0} \right),
\end{equation}
\begin{equation}\label{fp}
\frac{p}{\rho_0 u_0^2}=f_p\left(-\frac{t}{t_0}\right),
\end{equation}
\begin{equation}\label{ftheta}
\theta =\frac{m p}{2 \rho}=\frac{m {u_0}^2}{2}\frac{f_p\left(-\frac{t}{t_0} \right)}{f_\rho \left(-\frac{t}{t_0} \right)}=\frac{m {u_0}^2}{2}f_\theta \left(-\frac{t}{t_0} \right)
\end{equation}
where $m$ is the mean mass of ions, electron and ion temperature are assumed as equal, $\theta$ is written for $kT$.  Also $f_\theta =\frac{f_p}{f_\rho}$.  For evaluating the functions in (\ref{frho}),(\ref{fp}) and (\ref{ftheta}) for density, pressure and temperature the initial conditions can be taken as the jump conditions.  When the first converging shock is passed a discontinuity is produced at the center.  The computations of the functions $f_\rho,f_p$ and $f_\theta$ can be seen in Goldman \cite{goldman1973}.
The fusion energy can be evaluated by integrating the thermonuclear reaction rate depending on position and time via density and temperature.
\begin{equation}
E_{fusion}=4 \pi E_{ij}\int_{0}^{r_{max}}r^2 {\rm d} r \int_{t_0}^{t_{max}} r_{ij}(r,t){\rm d} t
\end{equation}
where $E_{ij}$ denotes the energy released by a fusion reaction between the nuclei $i$ and $j$.\\

Using the reaction rate $r_{ij}$ given in  (\ref{reactionrate}) we have
\begin{eqnarray}
r_{ij}&=&(1-\frac{1}{2}\delta_{ij})N_i N_j \langle\sigma v\rangle_{ij} \nonumber\\
&=&(1+\delta_{ij})^{-1} \frac{N^2}{4} \langle\sigma v\rangle_{ij}.
\end{eqnarray}
Here $N$ denotes the total ion density of the fusion plasma, $\delta_{ij}$ is the Kronecker delta.  The factor $\frac{1}{4}$ is due to the fact that each reactant density is half the total ion number density.  Now take $N=\frac{\rho}{m}$ where $N$ is the ion density  and $\rho$ is the local density.  The dimensionless variable $\tau=-\frac{t}{t_0}$ which is positive since $t_0<0$.  We have
\begin{equation}\label{fusionenergytheta}
E_{fusion}= \pi (1+\delta_{ij})^{-1} {N_0}^2 E_{ij}\int_{0}^{r_{max}}r^2 \left(\frac{r}{\xi}\right)^{\frac{1}{n}} \int_{-1}^{\tau_{max}} {f_\rho}^2(\tau) \langle\sigma v=F(\theta(r,\tau))\rangle_{ij}{\rm d} \tau {\rm d} r
\end{equation}
where $N_0= \frac{\rho_0}{m}, \tau_{max}=-\frac{t_{max}}{t_0}$.\\

From Goldman \cite{goldman1973} we can see that the passage of the first converging shock gives a density increase followed by an adiabatic compression to a density ratio.  Then the reflected shock at the center on returning gives a further shock compression to a maximum density \cite{brueckner1974}.  It should be noted that temperature is a rapidly varying function of $r$ but slowly varying function of $\tau$ in the reflected shock.  So in (\ref{fusionenergytheta}) we can write
\begin{equation}
\langle\sigma v=F(\theta(r,\tau)\rangle_{ij}=\langle\sigma v=F(\theta(r,\tau_1))\rangle_{ij}
\end{equation}
where $\tau_1$ is the scaled time of the reflected shock.  So in the integral with respect to $\tau$ we can take  $\tau_1=-\frac{t}{t_0}$ instead of $-1=-\frac{t_0}{t_0}$.
For smaller values of $r$ and near the time of maximum compression the fusion yield $E_{fusion}$ shows a rapid increase.  So we can set the upper limits of integration in (\ref{fusionenergytheta}) as $\infty$.  Now (\ref{fusionenergytheta}) can be modified as \begin{equation}\label{fusionenergymodified}
E_{fusion}= \pi (1+\delta_{ij})^{-1} {N_0}^2 E_{ij}\left(\frac{1}{\xi}\right)^{\frac{1}{n}}\int_{0}^{\infty}r^{2+\frac{1}{n}} \int_{\tau_1}^{\infty} {f_\rho}^2(\tau) \langle\sigma v=F(\theta(r,\tau))\rangle_{ij}{\rm d} \tau {\rm d} r
\end{equation}

In front of the secondary shock the relation connecting the density, temperature and pressure at a radius $r$ is given by
\begin{equation}\label{frhos}
\frac{\rho}{\rho_s}=f_\rho(\tau),
\end{equation}
\begin{equation}\label{fps}
\frac{p}{p_s}=f_p(\tau),
\end{equation}
\begin{equation}\label{fthetatau}
\theta =\frac{m p}{2 \rho}=\frac{m p_s}{2 \rho_s}\frac{f_p(\tau)}{f_\rho (\tau)}=\frac{m p_s}{2 \rho_s}f_\theta (\tau).
\end{equation}
The equations are from Haubold and John \cite{hauboldjohn1981}.  By the jump conditions across the shock \cite{guderley1942, goldman1973}
\begin{equation}
\rho_s=\rho_0 \frac {\gamma+1}{\gamma-1}, \gamma \neq \pm 1
\end{equation}
\begin{equation}
p_s=\frac{2 \rho_0 {u_0}^2}{\gamma+1}
\end{equation}
and using equation (\ref{shockvelocity}) the temperature following the secondary shock is a function of $r$ given by
\begin{eqnarray}\label{theta}
\theta(r,\tau_1)&=&\frac{m {u_0}^2}{\gamma+1}\frac{\gamma-1}{\gamma+1}f_\theta (\tau_1)\nonumber\\
&=&\frac{m n^2 \xi^{\frac{2}{n}} r^{\frac{2(n-1)}{n}} }{\gamma+1}\frac{\gamma-1}{\gamma+1}f_\theta (\tau_1).
\end{eqnarray}
From (\ref{theta}) the radical variable $r$ is obtained as
\begin{eqnarray}
r&=&\left[\frac{m n^2 \xi^{\frac{2}{n}} }{\gamma+1}\frac{\gamma-1}{\gamma+1}f_\theta (\tau_1)\right]^{-\frac{n}{2(n-1)}} \left[ \theta(r,\tau_1)\right]^{\frac{n}{2(n-1)}}\nonumber \\
&=&\lambda^{-\frac{n}{2(n-1)}} \left[ \theta(r,\tau_1)\right]^{\frac{n}{2(n-1)}},
\end{eqnarray}
where
\begin{equation}
\lambda=\frac{m n^2 \xi^{\frac{2}{n}} }{\gamma+1}\frac{\gamma-1}{\gamma+1}f_\theta (\tau_1)
\end{equation}
Changing the variable of integration from the radical variable $r$ to the thermal energy variable $\theta$, the equation (\ref{fusionenergymodified}) becomes
\begin{eqnarray}\label{fusionenergymodifiedtheta}
E_{fusion}&=& \pi (1+\delta_{ij})^{-1} {N_0}^2 E_{ij}\left(\frac{1}{\xi}\right)^{\frac{1}{n}}\int_{0}^{\infty} \frac{n \lambda^{\frac{3n+1}{2(1-n)}} }{2(1-n)} \theta^{\frac{n+3}{2(n-1)}} \int_{\tau_1}^{\infty} {f_\rho}^2(\tau) \langle\sigma v=F(\theta)\rangle_{ij}{\rm d} \tau {\rm d} \theta \nonumber \\
&=&\pi (1+\delta_{ij})^{-1} {N_0}^2 E_{ij}I_1 C_n \int_{0}^{\infty} \theta^{\frac{n+3}{2(n-1)}} \langle\sigma v=F(\theta)\rangle_{ij} {\rm d} \theta,
\end{eqnarray}
where
\begin{eqnarray}
I_1&=&\int_{\tau_1}^{\infty} {f_\rho}^2(\tau) {\rm d} \tau  \\
C_n&=&\frac{n}{2(1-n)} \lambda^{\frac{3n+1}{2(1-n)}} \left(\frac{1}{\xi}\right)^{\frac{1}{n}}.
\end{eqnarray}
Now we evaluate the fusion energy released by the plasma substituting the extended thermonuclear reaction probability over temperature.
\section{\bf Evaluation of the fusion energy integral by the extended thermonuclear reaction probability in closed form}
 With the help of the closed form evaluation representation of the reaction probability obtained in section 3 we find the closed form of the fusion energy yield integral (\ref{fusionenergymodifiedtheta}) in the previous section.  The calculations are done by taking the cross section as a non-resonant case in the energy region.  It is found in literature that in the laboratory experiments, for example reactions like $D(d,n)^3 He$ and $D(d,p)^3 H$, we take non-resonant case.  In equation (\ref{crosssectionfactor}) of section 2 if $S^{(\nu)}=0$ for $\nu=1$ and $\nu=2$.  Then we obtain the fusion yield integral after inserting the reaction probability (\ref{reactionprobabilitymathai}) in (\ref{fusionenergymodifiedtheta}) by writing $\theta$ for $kT$.
 \begin{eqnarray*}
 E_{fusion}&=& (1+\delta_{ij})^{-1} {N_0}^2 E_{ij}I_1 C_n S(0)\left( \frac{8}{\mu} \right)^{\frac{1}{2}} \frac{1}
{(\alpha-1 )\Gamma \left
(\frac{1}{\alpha-1 } \right )} \int_{0}^{\infty} \theta^{\frac{2}{n-1}}\nonumber \\
&&\times G_{1,3}^{3,1} \left[
(\alpha-1)\pi^2 \left( \frac{\mu}{2} \right)\left(\frac{Z_i Z_j e^2}{\hbar }\right)^2 \theta^{-1} \big|^{2-\frac{1}{\alpha-1}}_
 {1, \frac{1}{2}, 0} \right] {\rm d} \theta.
 \end{eqnarray*}
 By using the substitution $\theta=\frac{1}{u}$, we get

 \begin{eqnarray*}
 E_{fusion}&=& (1+\delta_{ij})^{-1} {N_0}^2 E_{ij}I_1 C_n S(0)\left( \frac{8}{\mu} \right)^{\frac{1}{2}} \frac{1}
{(\alpha-1 )\Gamma \left
(\frac{1}{\alpha-1 } \right )} \int_{0}^{\infty} u^{\frac{2n}{1-n}}\nonumber \\
&&\times G_{1,3}^{3,1} \left[
(\alpha-1)\pi^2 \left( \frac{\mu}{2} \right)\left(\frac{Z_i Z_j e^2}{\hbar }\right)^2 u \big|^{2-\frac{1}{\alpha-1}}_
 {1, \frac{1}{2}, 0} \right] {\rm d} u.
 \end{eqnarray*}
 By taking
 \begin{equation}
 \frac{(2\mu)^{\frac{1}{2}} \pi Z_i Z_j {\rm e}^2}{\hbar}=\bar{x}
  \end{equation}
  we get,
  \begin{eqnarray}\label{fusionyieldgfunction}
 E_{fusion}&=& (1+\delta_{ij})^{-1} {N_0}^2 E_{ij}I_1 C_n S(0)\left( \frac{8}{\mu} \right)^{\frac{1}{2}} \frac{1}
{(\alpha-1 )\Gamma \left
(\frac{1}{\alpha-1 } \right )} \int_{0}^{\infty} u^{\frac{n+1}{1-n}-1}\nonumber \\
&&\times G_{1,3}^{3,1} \left[
\frac{(\alpha-1)\bar{x}^2 }{4 }u \big|^{2-\frac{1}{\alpha-1}}_
 {1, \frac{1}{2}, 0} \right] {\rm d} u.
 \end{eqnarray}
In view of the Mellin transform of a $G$-function, equation (\ref{fusionyieldgfunction}) becomes
\begin{eqnarray}\label{fusionyieldgfunctionmodified}
 E_{fusion}&=& (1+\delta_{ij})^{-1} {N_0}^2 E_{ij}I_1 C_n S(0)\left( \frac{8}{\mu} \right)^{\frac{1}{2}} \frac{1}
{(\alpha-1 )\Gamma \left
(\frac{1}{\alpha-1 } \right )}\left[ \frac{(\alpha-1)\bar{x}^2 }{4 }\right]^{-\frac{n+1}{1-n}}\nonumber \\
&&\times \Gamma\left(1+\frac{n+1}{1-n}\right)\Gamma\left(\frac{1}{2}+\frac{n+1}{1-n}\right)
\Gamma\left(\frac{n+1}{1-n}\right)
\Gamma\left(\frac{1}{\alpha-1}-1-\frac{n+1}{1-n}\right)\nonumber\\
 \end{eqnarray}
 where $\alpha>1, \frac{1}{\alpha-1}-1-\frac{n+1}{1-n}>0$.\\
 By Legendre's Duplication formula \cite{mathai1993, mathaietal2010},
  \begin{equation}
  \Gamma(2z)=\pi^{-\frac{1}{2}}2^{2z-1}\Gamma(z)\Gamma\left(z+\frac{1}{2}\right)
 \end{equation}
 we get,
\begin{eqnarray}\label{fusionyieldmodified}
 E_{fusion}&=& (1+\delta_{ij})^{-1} {N_0}^2 E_{ij}I_1 C_n S(0)\left( \frac{8\pi}{\mu} \right)^{\frac{1}{2}} 2^{-\frac{1+3n}{1-n}}\frac{1}
{(\alpha-1 )\Gamma \left
(\frac{1}{\alpha-1 } \right )}\left[ \frac{(\alpha-1)\bar{x}^2 }{4 }\right]^{-\frac{n+1}{1-n}}\nonumber \\
&&\times \Gamma\left(\frac{2}{1-n}\right)\Gamma\left(\frac{2(1+n)}{1-n}\right)
\Gamma\left(\frac{1}{\alpha-1}-\frac{2}{1-n}\right), \frac{1}{\alpha-1}-\frac{2}{1-n}>0
 \end{eqnarray}
 where
 \begin{eqnarray*}
I_1&=&\int_{\tau_1}^{\infty} {f_\rho}^2(\tau) {\rm d} \tau  \\
C_n&=&\frac{n}{2(1-n)} \lambda^{\frac{3n+1}{2(1-n)}} \left(\frac{1}{\xi}\right)^{\frac{1}{n}}\\
&=&\frac{n^{\frac{2(1+n)}{1-n}}}{2(1-n)} \xi^{\frac{4}{1-n}}
\left[\frac{m }{\gamma+1}\frac{\gamma-1}{\gamma+1}f_\theta (\tau_1)\right]^{\frac{1+3n}{2(1-n)}}.
\end{eqnarray*}
For a physically realizable solution we take $n=0.68837$. Then equation (\ref{fusionyieldmodified}) becomes
\begin{eqnarray}\label{fusionyieldparticular}
 E_{fusion}&=& (1+\delta_{ij})^{-1} {N_0}^2 E_{ij}I_1 C_n S(0)\left( \frac{32\pi}{\mu} \right)^{\frac{1}{2}} \frac{1}{(\alpha-1 )\Gamma \left
(\frac{1}{\alpha-1 } \right )}\left[ (\alpha-1)\bar{x}^2 \right]^{-5.418}\nonumber \\
&&\times \Gamma(6.418)\Gamma(10.836)
\Gamma\left(\frac{1}{\alpha-1}-6.418\right)
 \end{eqnarray}
Now we compare the newly obtained result with the existing results in the next section.
\section{\bf Comparison of the results and interpretation of the pathway parameter}

The fusion yield integral in (\ref{fusionyieldparticular}) is obtained by using the extended thermonuclear reaction rate (\ref{reactionprobabilitymathai}) in (\ref{fusionenergymodifiedtheta}).  If we take the limit of $E_{fusion}$ as $\alpha \rightarrow 1$ we obtain the fusion yield integral in the Maxwell-Boltzmann case which is given in Haubold and John \cite{hauboldjohn1981}.
\begin{equation}\label{fusionyieldparticularstandard}
 E_{fusion}^{(1)}= (1+\delta_{ij})^{-1} {N_0}^2 E_{ij}I_1 C_n S(0)\left( \frac{32\pi}{\mu} \right)^{\frac{1}{2}} [\bar{x}^2] ^{-5.418} \Gamma(6.418)\Gamma(10.836),
 \end{equation}
 by using the asymptotic expansion of gamma function \cite{Erdelyi1953}
 \begin{equation}
 \Gamma(z+a)\sim (2 \pi)^\frac{1}{2} z^{z+a-\frac{1}{2}} e^{-z},z \rightarrow \infty, |arg(z+a)|<\pi-\epsilon, \epsilon>0
 \end{equation}
 where the symbol $\sim$ means asymptotically equivalent to.\\

   The interpretation of the pathway parameter used in extending the results can be done with the help of the $\delta^{th}$ moment.  By using the type-2 beta form of the pathway model given in (\ref{type2pathway}) and the normalizing constant in (\ref{constanttype2}), the $\delta^{th}$-moment in the case of $\alpha>1$ is
   \begin{eqnarray}\label{deltamoment}
\mathcal{E}(x^\delta)&=&
 \frac{\delta[a(\alpha-1)]^{\frac{\gamma+1}{\delta}}\Gamma\left( \frac{1}{\alpha-1}\right)}{
\Gamma\left( \frac{\gamma+1}{\delta}\right)\Gamma\left( \frac{1}{\alpha-1}-\frac{\gamma+1}{\delta}\right)}\int _{0}^{\infty}x^{\gamma +\delta }
{[1+a(\alpha-1 )x^\delta]}^{-\frac{1}{(\alpha-1 )}}{\rm d}x \nonumber \\
&=&\frac{\gamma+1}{a [(2 \delta+\gamma+1)-\alpha(\gamma+1+\delta)]},~x>0,\alpha>1.
\end{eqnarray}
where $\mathcal{E}(.)$ denotes the expected value of (.).  The value of $\alpha$ in terms of the $\delta^{th}$ moment is given by
\begin{equation}\label{valueofalpha}
\alpha=1+\frac{1}{(\gamma+1+\delta)}\left[ \delta-\frac{\gamma+1}{a \mathcal{E}(x^\delta)}\right],~x>0,\alpha>1.
\end{equation}
 For $\gamma>0, \delta>0, a>0$, the pathway parameter $\alpha$ increases and finally when the mean value or expected value of $x^\delta$ decreases $\alpha$ goes to $-\infty$.  While deriving the pathway model by optimizing Mathai's entropy one of the constraints was that $\mathcal{E}(x^\delta)$ is given.  The value of $\alpha$ is determined depending upon the preselected value of $\mathcal{E}(x^\delta)$.  When $\mathcal{E}(x^\delta)$ is set at high value then $\alpha$ approaches $1+\frac{\delta}{(\gamma+1+\delta)}$ and if $\mathcal{E}(x^\delta)$ is a lower number, near to zero, then $\alpha$ will approach $-\infty$.  When $\alpha \rightarrow 1$ we have $\mathcal{E}(x^\delta)=\frac{\gamma+1} {a \delta}$.  The values of $\mathcal{E}(x^\delta)$ and $\alpha$ in the case of $\alpha<1$ is similar to (\ref{deltamoment}) and (\ref{valueofalpha}).  Therefore
 \begin{equation*}
\alpha \lesseqqgtr 1 \Rightarrow \mathcal{E}(x^\delta) \lesseqqgtr \frac{\gamma+1}{a\eta \delta}
\end{equation*}

In (\ref{reactionprobabilitysubstituted}) we have taken $y=\frac{E}{kT}$.  Here if $T$ increases $y$ will decrease and vice versa.    If the expected energy $\mathcal{E}(\frac{E}{kT})<\frac{3}{2}$ then $\alpha<1$, if $\mathcal{E}(\frac{E}{kT})=\frac{3}{2}$ then $\alpha=1$ and when $\mathcal{E}(\frac{E}{kT})>\frac{3}{2}$ then $\alpha>1$.  The value of $\alpha$ in (\ref{mathaipathway}) is obtained by putting $\gamma=\frac{1}{2}, \delta=1$ and $a=1$ in (\ref{valueofalpha}) and is given by
\begin{equation*}
\alpha=\frac{7}{5}-\frac{3}{5\mathcal{E}\left( \frac{E}{kT}\right)}\Rightarrow\alpha\leq
 \frac{7}{5}~ \text{or}~ \alpha>-\infty.
\end{equation*}

\section{\bf Conclusion}

We extended the thermonuclear reaction rate from the Maxwell-Boltzmann approach to the Tsallis approach and obtained an analytical closed-form representation of the extended reaction rate probability integral.  Instead of using the Maxwell-Boltzmann density function in equation (\ref{Maxwelboltzman}) we used pathway the density function (\ref{mathaipathway}) and the extended reaction rate has been represented in terms of Meijer's $G$-function.  We also obtained a linear homogeneous differential equation (\ref{mathaidifferentialequation}) of order 3 satisfied by the $G$-function of the extended reaction rate.  Then by using the Guderley model for the hydrodynamical compression and heating of the fusion plasma by a single laser-driven spherical shock, a general formula for the fusion yield integral is obtained by using the closed form representation of the extended thermonuclear reaction rate.  The results obtained are compared with the result  by Haubold and John \cite{hauboldjohn1981}. An interpretation of the pathway parameter is also given.\\

\noindent
{\bf Acknowledgment}\\

The authors would like to thank the Department of Science and Technology, Government of India, New Delhi, for the financial assistance for this work under project No. SR/S4/MS:287/05, and the Centre for Mathematical Sciences
for providing all facilities. The authors are particularly grateful for advice from Professor A.M. Mathai, Director of the Centre for Mathematical Sciences.\\

{\small

 \end{document}